\documentclass[12pt,doublespace]{article}
\begin{document}
\begin{center}
ERASURE OF TIME DELAY SIGNATURES IN THE OUTPUT OF AN OPTOELECTRONIC FEEDBACK LASER WITH MODULATED DELAYS 
AND CHAOS SYNCHRONIZATION\\
E.M.Shahverdiev $^{1,2}$ and K.A.Shore $^{1}$\\
$^{1}$School of Electronics,University of Wales, Bangor, Dean St.,Bangor, LL57 
1UT, Wales, UK\\
$^{2}$Institute of Physics, H.Javid Avenue,33, Baku, AZ1143, Azerbaijan\\
ABSTRACT
\end{center}
By studying the autocorrelation function of the optoelectronic feedback semiconductor laser output we establish that the signatures of time delays can be erased in systems incorporating modulated feedback time delays. This property is of importance for the suitability of such laser systems for secure chaos-based communication systems. We also make the first report on chaos synchronization in both unidirectionally and bidirectionally coupled multiple time delay chaotic semiconductor lasers with modulated optoelectronic feedbacks.\\
\begin{center}
I. INTRODUCTION
\end{center}
\indent Chaos synchronization [1] is a significant phenomenon in nonlinear science and plays an important role in a variety of complex physical, chemical, biological, power etc. systems, see e.g.references in [2]. Particularly,chaos synchronization is of immense importance to secure communications based on chaos control.In a chaos based secure communications a message is masked in the broadband chaotic output of the transmitter laser and 
synchronization between the transmitter and receiver systems is used to recover a transmitted message. Semiconductor lasers which are an integral part of high speed optical communication systems are also widely used in studies chaos-based communications, as they are readily available and can easily be driven to chaos with feedback [3-4].\\
\indent In coherent optical feedback laser chaos-based communication systems synchronization performance sensitively
depends on the detuning between the frequencies of the transmitter and reciever lasers. In particular detuning by a few hundred megahertz between the lasers could lead to a large degradation of the synchronization, which prevents message extraction at the receiver laser. Then from the practical point of view it is of great interest to employ laser systems which do not require fine tuning of the lasers frequencies. Semiconductor lasers subject to opto-electronic feedbacks are one such system. Compared with optical feedback, opto-electronic feedback is flexible and reliable  because of its convenience to be electrically controlled and its insensitivity to optical phase variations. Moreover, optoelectronic coupling avoids the complexity introduced by the phase of the electric field that otherwise plays a crucial role when considering the optical coupling, see, e.g.[5] and references there-in.Routes to optical chaos in semiconductor lasers with delayed optoelectronic feedback is reported in [6].\\
\indent Laser systems with multiple time delays could be of particular interest for secure chaos-based communication.
Recently it has been shown [7] that the incorporation of  additional time delays can  play a  stabilizing role in chaos synchronization  and also can offer a higher complexity of dynamics than achievable in more conventional single delay time systems. \\
\indent Identification of time delays of such systems can allow the eavesdropper to extract the message successfully using a simple local reconstruction of the time delay system. The delay time can be revealed by a number of means including via the autocorrelation coefficient [8].\\
\indent In this paper we investigate the autocorrelation function of the optoelectronic feedback laser output and establish that variable time delays result in the loss of signatures of time delays. In this way it is established that variable time delay systems can offer an enhanced robustness to eavesdropper attack compared to the case of fixed time-delay systems. The fundamental requirement for enabling message extraction in a chaos-based communication is the ability to achieve high-quality synchronization. In this respect this paper also presents the first report of the chaos synchronization properties of  both uni-directionally and bi-directionally coupled variable multiple time delayed semiconductor lasers with two optoelectronic feedbacks.\\
\begin{center}
II. SYSTEM MODEL
\end{center}
\indent Consideration is given to a system composed by two coupled identical single mode semiconductor lasers subject to optoelectronic feedbacks. In a master/slave configuration the optical power emitted by each laser is divided into several parts, detected, amplified, and added to their own injection current;also the optical power emitted by the master laser is detected, amplified, and added to the bias current of the slave laser (figure 1) Thus the dynamics of the double time delay master/transmitter laser is governed by the following systems
\begin{equation}
\frac{dS_{1}}{dt}=(\Gamma g_{1} -\gamma_{c})S_{1}
\end{equation}
\begin{equation}
\frac{dN_{1}}{dt}=I_{1} - \gamma_{s1} N_{1} - g_{1}S_{1} + \gamma_{c}(k_{1} S_{1}(t-\tau_{1}) + k_{2} S_{1}(t-\tau_{2}) + K_{1}S_{2}(t-\tau_{3}))
\end{equation}
The slave/receiver laser is described by  
\begin{equation}
\frac{dS_{2}}{dt}=(\Gamma g_{2} -\gamma_{c})S_{2}
\end{equation}
\begin{equation}
\frac{dN_{2}}{dt}=I_{2} - \gamma_{s2} N_{2} - g_{2}S_{2} + \gamma_{c}(k_{3} S_{2}(t-\tau_{1}) + k_{4} S_{2}(t-\tau_{2}) + KS_{1}(t-\tau_{3})) 
\end{equation}
where subindicies 1,2 distinguish between the transmitter and receiver;$S_{1,2}$ is the photon density; $N_{1,2}$ is the carrier density;
$g_{1,2}$ is the material gain;$\gamma_{c}$ is the cavity decay rate;$\gamma_{s}$ is the carrier relaxation rate;$\Gamma$ is the confinement factor of the laser waveguide.$I_{1,2}$ is the injection current (in units of the electron charge);
$k_{1,2}$ and $k_{3,4}$ are the feedback rates for the transmitter and receiver systems,respectively; $\tau_{1,2}$
are the feedback delay times in the transmitter and receiver systems;$K$ is the coupling rate between the
transmitter and the receiver.$\tau_{3}$ is the time of flight between lasers. Term $K_{1}S_{2}(t-\tau_{3})$ in the right-hand side of Eq.(2)is valid only for mutually coupled systems. In a wide operation range the material gain $g$ can be expanded as 
\begin{equation}
g\approx g_{0} + g_{n}(N-N_{0}) + g_{p}(S-S_{0}),
\end{equation}
where $g_{0}=\gamma_{c}/\Gamma $ is the material gain at the solitary threshold; $g_{n}=\partial g/\partial N >0$ is the differential gain parameter; 
$g_{p}=\partial g/\partial S <0$ is the nonlinear gain parameter; $N_{0}$ is the carrier density at threshold; $S_{0}$ is the free-running intracavity photon density when the lasers are decoupled;the parameters $g_{n}$ and $g_{p}$ are taken to be approximately constant. It is noted that for $k_{1}=k_{3}=0$ (or $k_{2}=k_{4}=0$) we obtain the case of coupled laser systems with a single optoelectronic feedback.\\
In figure 1 a schematic diagram of the experimental set-up for synchronization is given. In the experimental scheme unidirectional coupling can be realized by inclusion of an optical isolator (OI) as shown in the figure.\\ 
In the paper we will mainly consider the variable time delays $\tau_{1}(t),\tau_{2}(t)$, and $\tau_{3}(t);$
we choose the following form for the modulation of time delays:
\begin{equation}
\tau_{1,2}=\tau_{01,02} + S^{f}_{1}(t) \tau_{a1,a2}\sin(\omega_{1,2}t)
\end{equation}
are the variable feedback delay times.  
\begin{equation}
\tau_{3}=\tau_{03} + S^{f}_{1}(t) \tau_{a3}\sin(\omega_{3}t)
\end{equation}
is the variable time of flight between the transmitter laser and receiver laser; $\tau_{01,02,03}$ are mean time delays, $\tau_{a1,a2}$ are the amplitude,$\omega_{1,2,3}/2\pi$ are the frequency of the modulations; $S^{f}_{1}(t)$ is the output power of the master laser, Eqs.(1-2)($K_{1}=0$) for constant time delays.\\
\indent In the case of variable time delays, establishing the existence and stability conditions for the synchronization is not as straightforward as for the fixed time delays. Having in mind that for $\omega_{1,2,3}=0$ we obtain a case of fixed time delays, then as an initial estimate one can benefit from the existence conditions for the fixed time delays case [5]. Comparing  Eqs. (1-2) and Eqs.(3-4) for the case of fixed time delays one finds that a synchronous solution
\begin{equation}
S_{2}(t) = S_{1}(t) ,     N_{2}(t) = N_{1}(t)                                                                                                               
\end{equation}
exists if
\begin{equation}
k_{1}=k_{3} + K,  k_{2}=k_{4}, \tau_{1}=\tau_{3}                                                                                                         
\end{equation}
The synchronous solution (8) also  exist if
\begin{equation}
k_{1}=k_{3},  k_{2}=k_{4} + K, \tau_{2}=\tau_{3}                                                                                                         
\end{equation}
In the case of mutually coupled systems the synchronous solution (8) exists under the conditions 
\begin{equation}
k_{1}=k_{3},  k_{2}=k_{4}, K_{1}=K                                                                                                         
\end{equation}
It is our conjecture that also in the case of variable time delays systems, high quality complete synchronization  $S_{2}(t) = S_{1}(t), N_{2}(t) = N_{1}(t)$ will be obtained if the parameters satisfy the conditions: 
$k_{1}=k_{3} + K,  k_{2}=k_{4},$  if $\tau_{1}(t) = \tau_{3}(t).$ Analogously complete synchronization $S_{2}(t) = S_{1}(t), N_{2}(t) = N_{1}(t)$ will also be established for $k_{2}=k_{4} + K,  k_{1}=k_{3},$  if $\tau_{2}(t) = \tau_{3}(t).$ The numerical simulations  described below demonstrate that this conjecture is well-founded.\\
\begin{center}
III. NUMERICAL SIMULATIONS AND DISCUSSIONS
\end{center}
In the numerical simulations we use typical values for the internal parameters of the transmitter laser:
$\gamma_{c}=2 ps,\gamma_{s1}=2 ns, \Gamma =0.3, g_{0}=7\times 10^{13},g_{n}=10^4,g_{p}=10^4, N_{0}=1.7\times 10^{8},S_{0}=5\times 10^{6}, 
I_{1}=3.4\times 10^{17} [5].$ The parameters at the receiver are chosen to be identical to those of the transmitter,except for the feedback levels.\\
\indent Before studying the synchronization between the laser systems with variable time delays we investigate the autocorrelation coefficient for the output of the master laser for both constant and variable time delays. The autocorrelation coefficient is a measure of how well a signal matches a time shifted version of itself and is a special case of the cross-correlation coefficient [9]
\begin{equation}
C(\Delta t)= \frac{<(x(t) - <x>)(y(t+\Delta t) - <y>)>}{\sqrt{<(x(t) - <x>)^2><(y(t+ \Delta t) - <y>)^2>}},
\end{equation}
where x and y  are the outputs of the interacting laser systems; the brackets $<.>$  represent the time average;  $\Delta t$  is a time shift between laser outputs. This coefficient indicates the quality of synchronization: C=0 implies no synchronization; C=1 means perfect synchronization.\\
Figure 2 demonstrates the autocorrelation coefficient (here the autocorrelation coefficient will be denoted by $C_{A}$ and obtained from equation (12) 
when $x=y$ ) for the output of the master laser, Eqs.(1-2)($K_{1}=0$) for constant time delays, i.e. for $\omega_{1,2}=0$ and with  $\tau_{01}=4ns, \tau_{02}=7ns, k_{1}=0.80, k_{2}=0.70$. It is clearly seen that time delays can be easily recovered from the autocorrelation coefficient, as it exhibits extrema at values of the time delays or their multiples and combinations.\\
\indent Next we consider the case of variable time delays. In investigating the behaviour of the autocorrelation coefficients we have experimented with different types of variable time delays, among them:(a)sinusoidal modulations $\tau_{1,2}=\tau_{01,02} + \tau_{a1,a2}\sin(\omega_{1,2}t)$; (b)chaotic modulations $\tau_{1,2}=\tau_{01,02} + S^{f}_{1}(t)$ and 
(c)product of chaotic and sinusoidal modulations $\tau_{1,2}=\tau_{01,02} + S^{f}_{1}(t) \tau_{a1,a2}\sin(\omega_{1,2}t).$\\
Extensive numerical simulations investigating the impact of (a), (b), and (c) type modulations on the autocorrelation function have established that the amplitude of the peaks and dips at the delays and their combinations are decreased significantly or eliminated totally more prominently for the type of modulation involving the product of sinusoidal and chaotic time delays modulations. Figure 3 depicts the autocorrelation coefficient of the laser output for the product of sinusoidal and chaotic time delays, Eqs.(1-2)($K_{1}=0$) for 
$\tau_{1}(t)= (4X10^{-9}  +  4X10^{-17}S^{f}_{1}(t)\sin(3X10^{6}t))$s and $\tau_{2}(t)= (7X10^{-9} + 4X10^{-17}S^{f}_{1}(t)\sin(3X10^{6}t))$s
with the rest of parameters as for figure 2. Thus, modulation of the delay times gives rise to the loss of their signature in the autocorrelation coefficient and therefore can improve the security of chaos based communication systems.\\
\indent As mentioned above, in chaos based communication schemes synchronization between the transmitter and receiver lasers is essential for successful message decoding. In order to establish the viability of the present approach we also present here the first report of chaos synchronization between variable time-delay lasers. Unidirectionally coupled laser systems, Eqs.(1-4) ($K_{1}=0$) were simulated for variable feedback time delays $\tau_{1}(t)= (4X10^{-9}  +  4X10^{-17}S^{f}_{1}(t)\sin(3X10^{6}t))$s and $\tau_{2}(t)= (7X10^{-9} + 4X10^{-17}S^{f}_{1}(t)\sin(3X10^{6}t))$s and  variable coupling time delay $\tau_{3}(t)= (4X10^{-9} + 4X10^{-17}S^{f}_{1}(t)\sin(3X10^{6}t))$s with parameter values $k_{1}=0.80, k_{2}=k_{4}=0.70, k_{3}=0.15, K=0.65$. Figure 4(a) portrays dynamics of the master laser intensity. Numerical simulations revealed that the dynamics of the slave laser follows that of the master laser exactly with the correlation coefficient C=1 (figure 4(b)). We note  that, although not presented here, the complete chaos synchronization regime was also observed for $ k_{2}=k_{4} + K,  k_{1}=k_{3},$ and $\tau_{2}(t) = \tau_{3}(t).$ That is, duality of the synchronization regimes established for constant time delay cases [5] is also valid for the case of variable time delays. \\
Figure 5(a)depicts the time series of the laser intensity $S_{1},$ Eqs.(1-4) for the case of mutually coupled lasers with the same variable feedback and coupling time delays as in figure 4 for parameters $k_{1}=k_{3}= 0.80, k_{2}=k_{4}=0.70, K=K_{1}=0.65.$  
Dynamics of $S_{2}$ is correlated with that of $S_{1}$ with the correlation coefficent C=1 (figure 5(b)). The values of the cross-correlation coefficients for both uni-directionally and bi-directionally coupled systems testify to the high quality chaos synchronization and thus demonstrate the efficacy of this approach to chaos-based communications.\\
According to Kerckhoffs' principle, the security of a cryptosystem can not be based on the secrecy of its encryption and decryption procedures. The security of the cryptosystem must be only related to the difficulty of guessing the key, and it can not  depend on the lack of knowledge about inner operating of the encryption and decryption procedures [10-11]. Recently in [12] the feedback time delay was proposed as a possible key for increased security of chaotic optical communication. Chaotic modulation of the feedback time delays can further enhance the security of such communication schemes. \\
Finally we emphasize that this paper deals with the ideal case, where the feedback and coupling loops, have a an infinite bandwidth. In the realistic case there are certain bandwidth requirements for the detection devices, etc. to see the fine structure of the laser dynamics.It should also be emphasized that numerical simulations revealed that synchronization is very robust to parameter mismatches (2-3$\%$), which is of certain importance from the practical point of view. Also,currently we are not aware  of the experimental realization of chaos synchronization in optoelectronic feedback semiconductor lasers with variable time delays. However given the considerable experience in the laboratory demonstrations of laser chaos synchronization with the use of off-the shelf devices, this task would not be particularly difficult. 
Moreover a recent field experiment for laser chaos-based communications demonstrated the viability of the chaos communication over an installed optical fibre network [13]. Particularly, dispersion was not found to have a major impact on the overall performance of the system in the field trial.
\begin{center}
IV.CONCLUSIONS
\end{center}
\indent To conclude, we have demonstrated that the time-delay signature is eliminated from the laser output autocorrelation in systems with modulated optoelectronic feedbacks.We have also described chaos synchronization in both uni-directionally and bi-directionally coupled variable multiple time delay lasers with optoelectronic feedbacks. The results of the paper provide the basis for the use of the optoelectronic feedback lasers with multiple variable time delays in enhanced security chaos-based high-speed communication systems.\\
\begin{center}
V.ACKNOWLEDGEMENTS
\end{center}
This research was supported by a Marie Curie Action within the 6th European Community Framework Programme Contract.\\
\newpage
\begin{center}
Figure captions
\end{center}
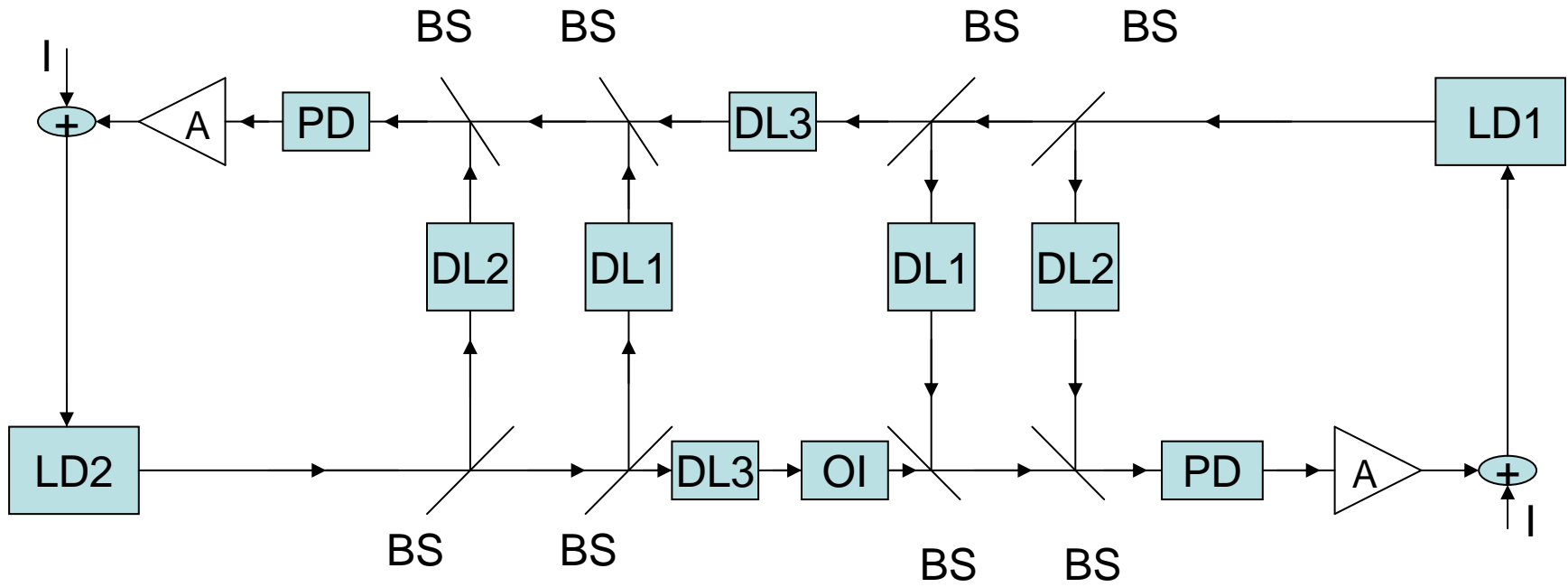
FIG.1.Schematic experimental arangement for the synchronization of lasers with double optoelectronic feedback and injection.LD:Laser diode. PD:Photodetector.BS:Beamsplitter. DL:Delay lines. OI:Optical Isolator. A:Amplifier. The output from each laser is split by a beamsplitters and directed along different feedback loops and coupling loops. Each signal is converted into  an electronic signal by a photodetector and then amplified before being added to the injection current of the laser.\\
~\\
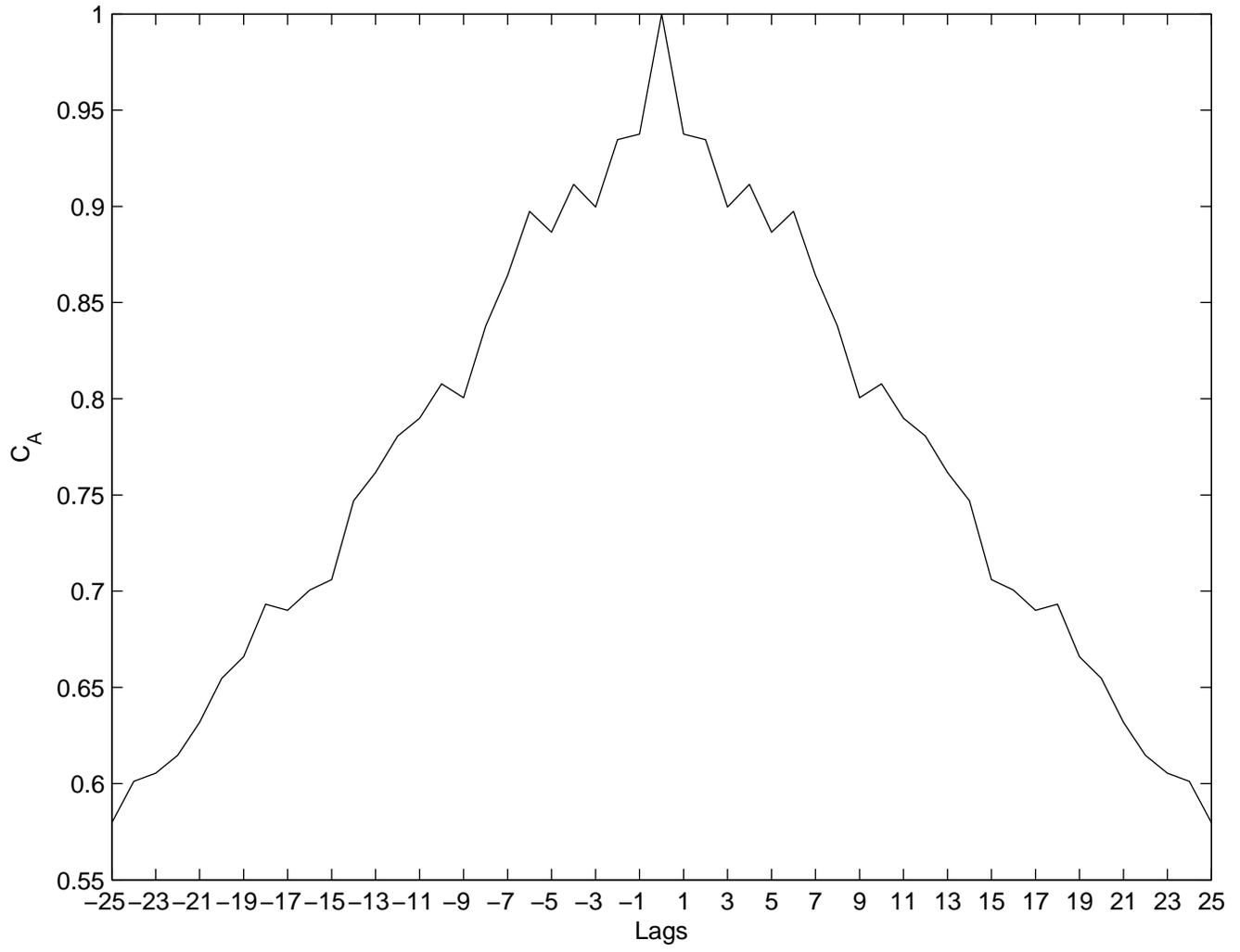
FIG.2. The autocorrelation coefficient $C_{A}$ of the laser output for constant time delays, Eqs.(1-2)($K_{1}=0$) for 
$\tau_{01}=4ns, \tau_{02}=7ns,  k_{1}=0.80, k_{2}=0.70.$ Lags are in ns.\\
~\\
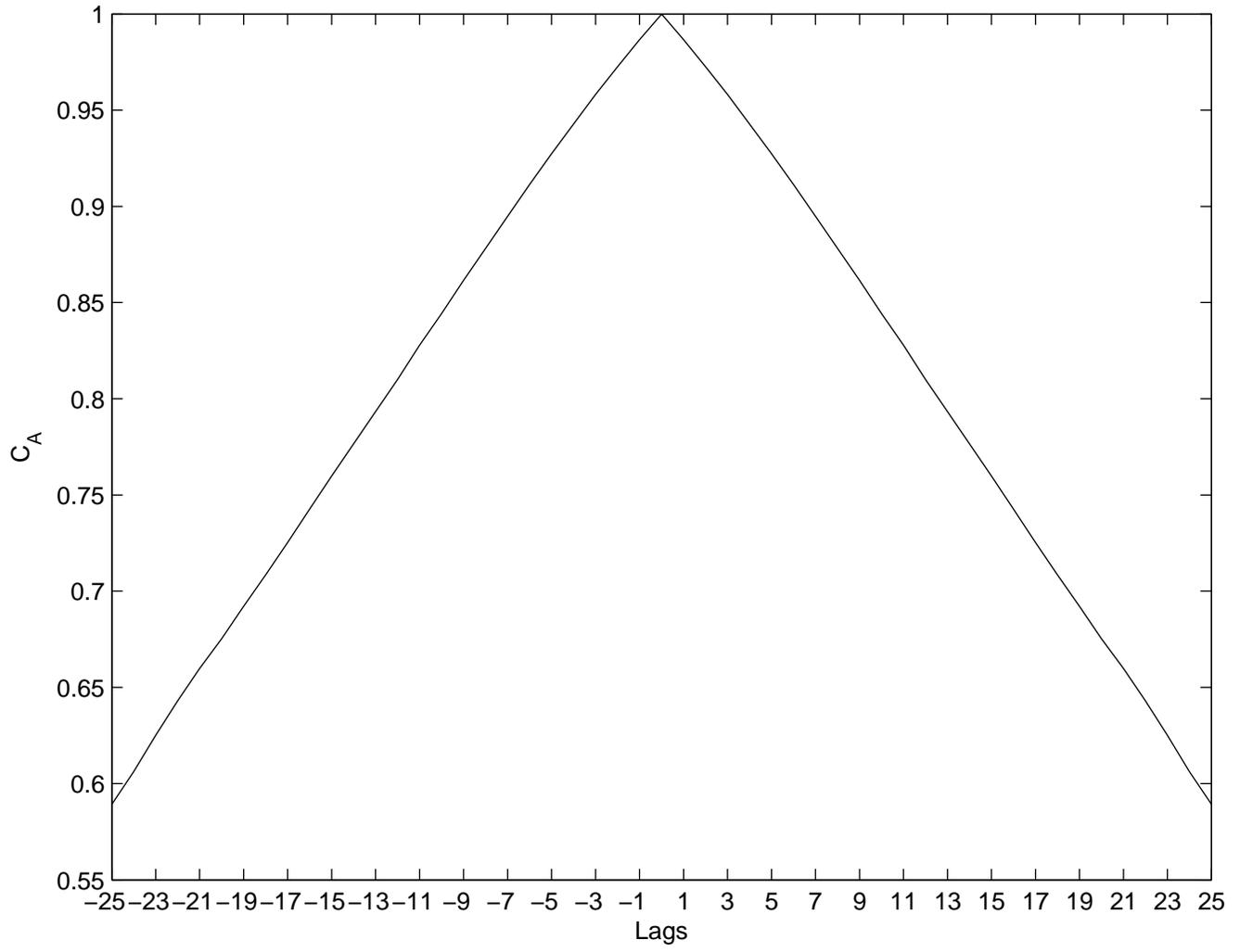
FIG.3. The autocorrelation coefficient $C_{A}$ of the laser output for the product of the sinusoidal and chaotic modulations of time delays, Eqs.(1-2)($K_{1}=0$) for $\tau_{1}(t)= (4X10^{-9}  +  4X10^{-17}S^{f}_{1}(t)\sin(3X10^{6}t))$s and $\tau_{2}(t)= (7X10^{-9}  +  4X10^{-17}S^{f}_{1}(t)\sin(3X10^{6}t))s.$ The other parameters are as in figure 2. Lags are in ns.\\
~\\
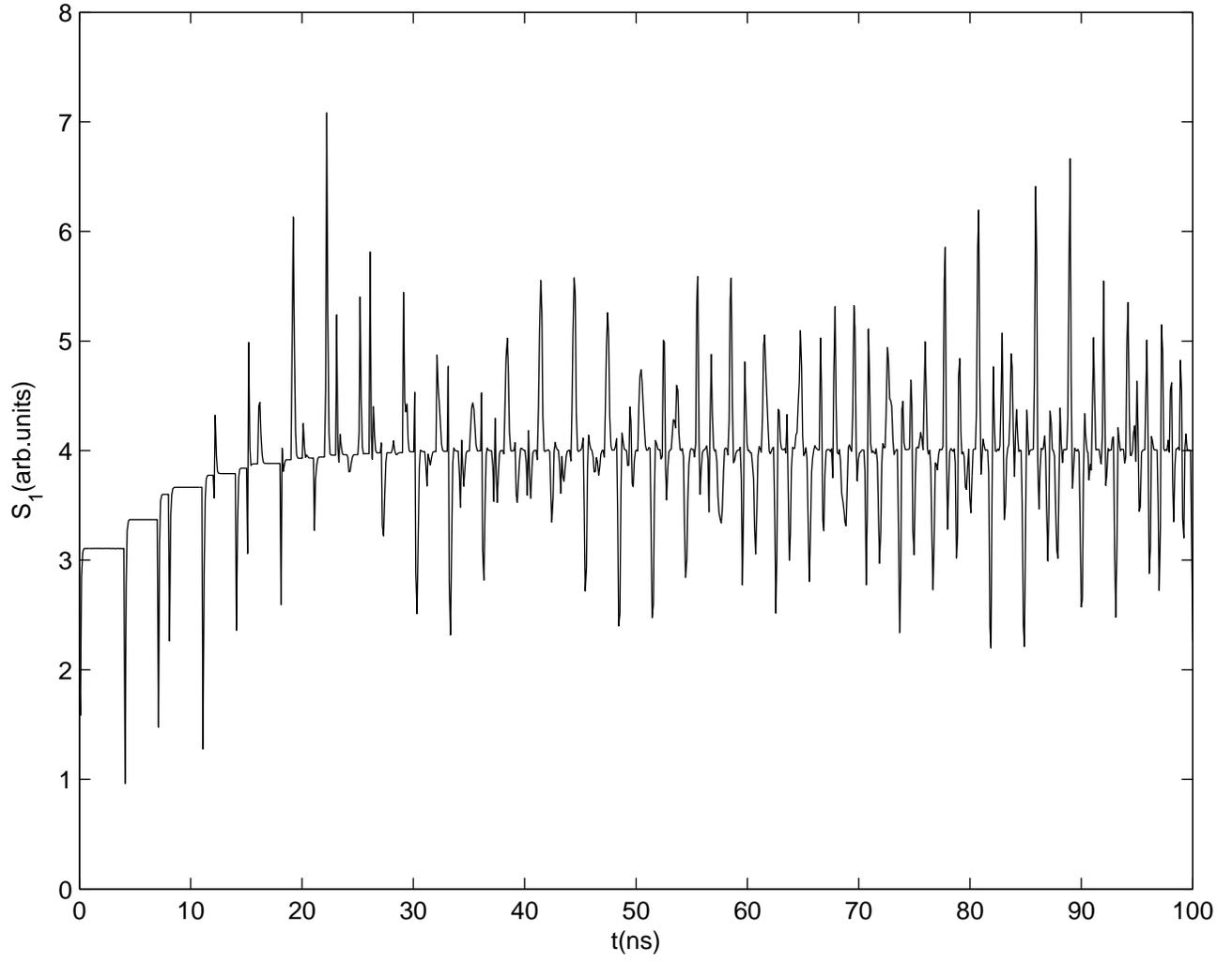
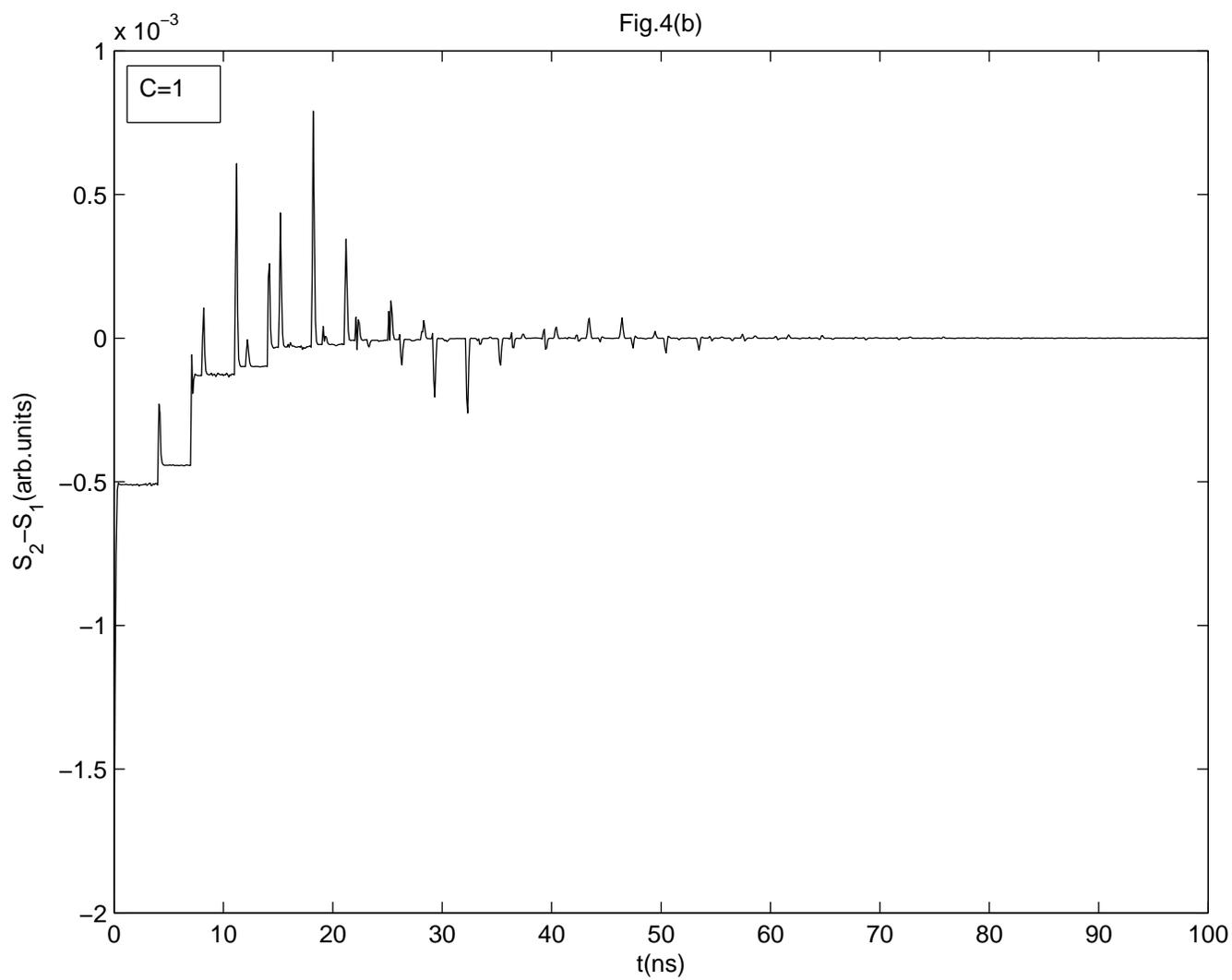
FIG.4. Numerical simulation of uni-directionally coupled variable time delay lasers, Eqs.(1-4)($K_{1}=0$) for 
$\tau_{1}(t)= (4X10^{-9}  +  4X10^{-17}S^{f}_{1}(t)\sin(3X10^{6}t))$s and \\
$\tau_{2}(t)= (7X10^{-9}  +  4X10^{-17}S^{f}_{1}(t)\sin(3X10^{6}t))s,$
$\tau_{3}(t)= (4X10^{-9}  +  4X10^{-17}S^{f}_{1}(t)\sin(3X10^{6}t))s, k_{1}=0.80, k_{2}=k_{4}=0.15, k_{3}=0.15, K=0.65.$ Complete synchronization: (a) time series of the transmitter laser intensity $S_{1}$;(b)Error $S_{2}-S_{1}$ dynamics. C is the cross-correlation coefficient between the intensities of the transmitter and receiver lasers.\\
~\\
FIG.7. Numerical simulation of bi-directionally coupled variable time delays lasers, Eqs.(1-4) for  the  parameters $\tau_{1}(t)= (4X10^{-9}  +  4X10^{-17} S^{f}_{1}(t)\sin(3X10^{6}t))$s and $\tau_{2}(t)= (7X10^{-9}  +  4X10^{-17}S^{f}_{1}(t)\sin(3X10^{6}t))s, \tau_{3}(t)= (4X10^{-9}  +  4X10^{-17}S^{f}_{1}(t)\sin(3X10^{6}t))s,k_{1}= k_{3}=0.80, k_{2}=k_{4}=0.70, K=K_{1}=0.65.$ Complete synchronization: (a) time series of the laser intensity $S_{1}$; (b) 
$S_{2}$ versus $S_{1}$. C is the cross-correlation coefficient between the intensities of the mutually coupled lasers.\\

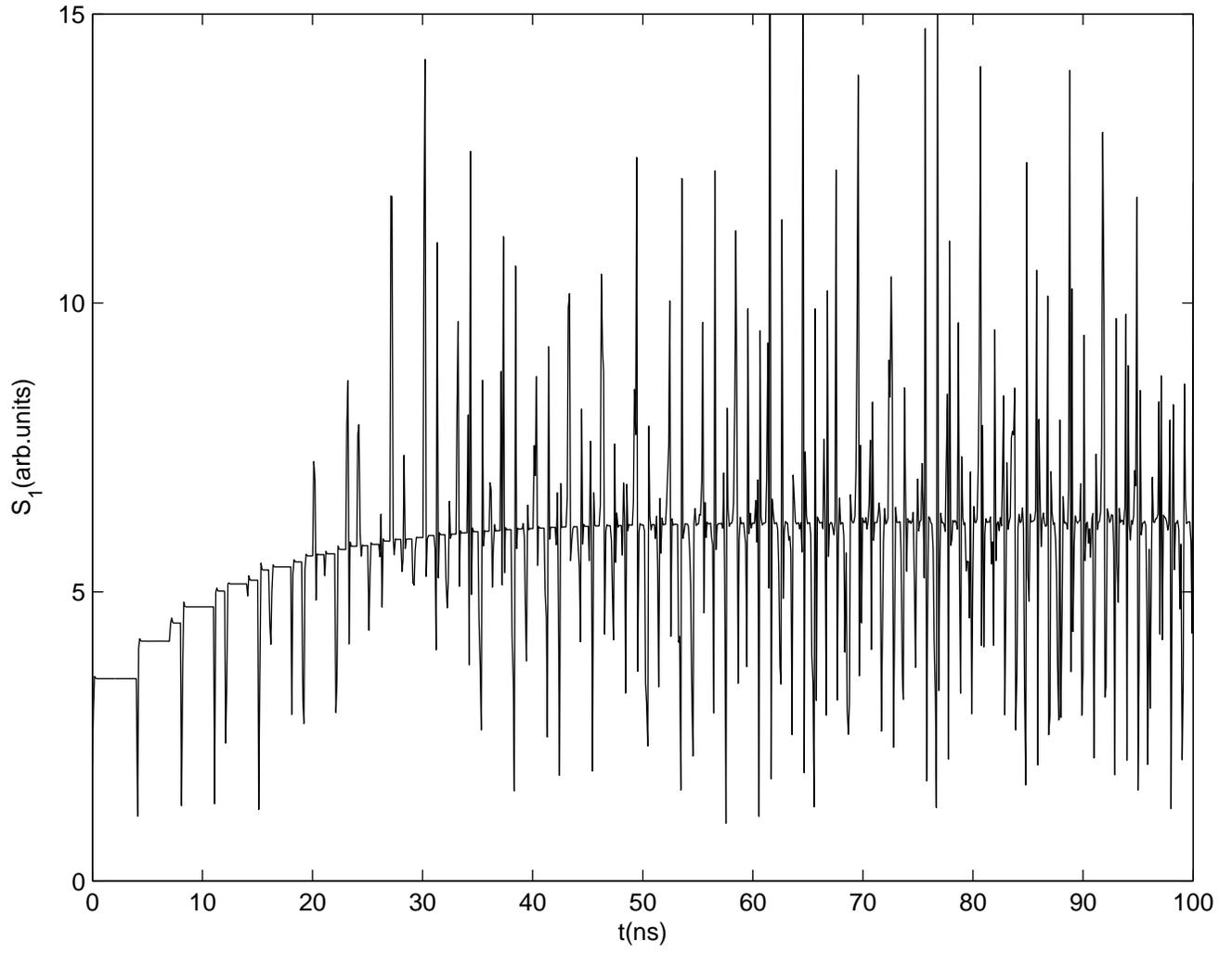
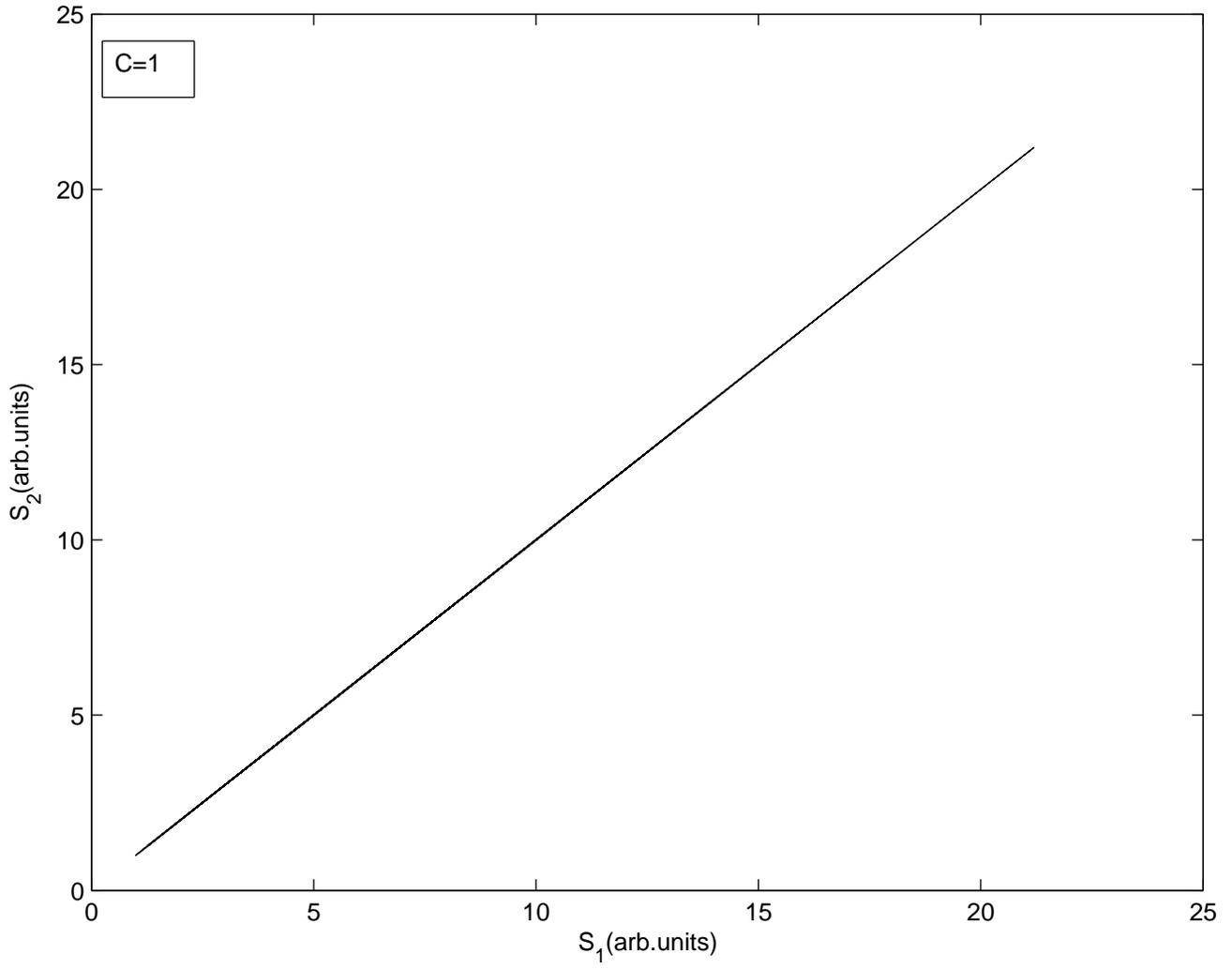
\end{document}